\documentclass[structabstract]{aa}

\usepackage{graphicx}
\usepackage{natbib}
\bibpunct{(}{)}{;}{a}{}{,} 
\usepackage{amsmath}
\usepackage{url}
\usepackage{txfonts}
\usepackage{multirow}
\usepackage[normalem]{ulem}

\begin{document}

\title{The extinction law from photometric data: linear regression
  methods\thanks{Based on observations collected at the European
    Organisation for Astronomical Research in the Southern Hemisphere,
    Chile (ESO programmes 069.C-0426 and 074.C-0728.)}}

\author{Joana Ascenso\inst1$^,$\inst2$^,$\inst{3} \and Marco Lombardi\inst{4} \and
  Charles J. Lada\inst{2} \and Jo\~ao Alves\inst{5}}
\institute{European Southern Observatory, Karl-Schwarzschild-Str. 2, 85748 Garching bei M\"unchen, Germany
\and Harvard-Smithsonian Center for Astrophysics,
  60 Garden Street, Cambridge, MA 02138, USA
\and Centro de Astrofisica da Universidade do Porto, Rua das Estrelas,
4150-762, Porto, Portugal
\and University of Milan, Department of Physics, via Celoria 16, 20133
Milan, Italy
\and University of Vienna, T\"urkenschanzstrasse 17, 1180 Vienna,
Austria}

\date{}

\abstract
{The properties of dust grains, in particular their size distribution,
  are expected to differ from the interstellar medium to the
  high-density regions within molecular clouds. Since the extinction
  at near-infrared wavelengths is caused by dust, the extinction law
  in cores should depart from that found in low-density environments
  if the dust grains have different properties.}
{We explore methods to measure the near-infrared extinction law
  produced by dense material in molecular cloud cores from photometric
  data. }
{Using controlled sets of synthetic and semi-synthetic data, we test
  several methods for linear regression applied to the specific
  problem of deriving the extinction law from photometric data. We
  cover the parameter space appropriate to this type of observations.}
{We find that many of the common linear-regression methods produce
  biased results when applied to the extinction law from photometric
  colors. We propose and validate a new method, LinES, as the most
  reliable for this effect. We explore the use of this method to
  detect whether or not the extinction law of a given reddened
  population has a break at some value of extinction.}
   {}


\titlerunning{The extinction law from photometric data: methods}
\authorrunning{J. Ascenso et al.}

\maketitle


\section{Introduction} \label{sec:introduction}

The properties of interestellar dust appear to be fairly constant
throughout the interstellar medium (ISM) of the Galaxy
\citep{RiekeLebofsky85,Kenyon:1998mz,Lombardi:2006gf,Jones:1980fr,Martin:1990zr},
reflecting the homogeneous physical conditions that characterize
it. In the cold molecular cores, however, under lower temperatures and
higher densities, the dust grains are believed to change, namely grow
by coalescence and/or develop ice mantles \citep[e.g.,
][]{Whittet88,Ossenkopf93,Whittet01,Draine03,Roman-Zuniga07,Steinacker10}. Measuring
these differences using methods other than detailed spectral analysis
has proved somewhat challenging, but the advent of larger and more
sensitive telescopes has started to reveal extinction laws toward
these regions that depart from the ISM typical curves, particularly in
the near- and mid-infrared regime\footnote{We will refer to
  near-infrared as the wavelength regime from 1 to 2.5 $\mu$m, and the
  mid-infrared from 3 to 8 $\mu$m.}, putting forward the extinction
law as a good indicator of grain properties.

Whereas the extinction law in low-density regions and the ISM is well
characterized by a power-law ($A_\lambda \propto \lambda^{-\beta}$) of
index $\beta\sim1.8$, several authors have found a pronounced
flattening of the extinction law in high density
regions. \citet{Lutz:1996aa} and \citet{Lutz:1999aa} first noted a
flat extinction law toward the Galactic center in this wavelength
range using spectroscopy of hydrogen recombination
lines. \citet{Nishiyama:2006ve} confirmed a flat extinction law toward
the Galactic center using the colors of red clump stars, later
confirmed independently by \citet{Fritz11}. Another example was found
by \citet{Indebetouw05}, using a different method based on {\it
  Spitzer} photometry, in the direction of and around the star forming
region RCW 49. A number of other studies on the extinction law using
{\it Spitzer} followed, namely \citet{Flaherty:2007aa},
\citet{Chapman:2009ab} and \citet{McClure:2009aa}, who found a gray
extinction law for star forming regions and molecular clouds;
\citet{Chapman:2009aa} and \citet{Roman-Zuniga07}, who found a gray
extinction law for cloud cores; and \citet{Nishiyama:2009aa}, who
found a similar law for the Galactic center. \citet{Chapman:2009aa},
\citet{Chapman:2009ab} and \citet{McClure:2009aa} go a step further by
analyzing the dependence of the extinction law with extinction,
finding that the extinction law becomes grayer at higher extinction
regimes. More recently, \citet{Cambresy11} measured an actual change
of the extinction law within the same region for a threshold of
$A_V=20$ mag in the Trifid Nebula.

In this paper we address the issue of determining the extinction law
from photometric data alone and the biases inherent to some of the
fitting methods used frequently in the literature. We begin by
defining the mathematical problem and the possible methods to extract
a linear fit (Sect. \ref{sec:extlaw}), validating each method with
synthetic data (Sec. \ref{sec:tests}). In
Sect. \ref{sec:detect-break-extinct} we address the issue of detecting
a flattening of the extinction law with extinction.

This is the first paper in a series of two. The following paper
(Ascenso et al., {\it in preparation}) will apply the results reported
here to actual observations of cores in the Pipe Nebula.

\section{The extinction law from photometric data} 
\label{sec:extlaw}

\subsection{Defining the problem}
\label{sec:problem}

The problem of deriving the extinction law from photometric data is
one of linear regression: the goal is to determine the slope of the
line that best fits the reddening-displaced positions of the stars in
a color-color diagram. This slope, $\beta$, is the ratio of two color
excesses that compose the color-color diagram, e.g.,
$\beta=E_{\lambda-K}/E_\mathit{H-K}$ in a $\lambda-K$ {\it vs.} $H-K$
diagram, and is, in this case, related to the extinction law
$A_\lambda/A_K$ by:

\begin{equation}
  \label{eq:9}
  \frac{A_\lambda}{A_K}=\left(\frac{A_H}{A_K}-1\right)\frac{E_{\lambda-K}}{E_\mathit{H-K}}+1
\end{equation}

Linear regression, however, is not a simple science. The presence of
errors in both coordinates, that they vary as a function of the
quantities being analyzed (heteroscedasticity), that they may be
correlated, and the presence of intrinsic scatter, make the required
linear regression analysis far more complex than the typical
chi-squared minimization of residuals.  Problems of this nature, in
particular applied to astronomical analysis, have been debated in the
literature for over 50 years \citep{Seares:1944aa,Trumpler:1953aa},
although every specific case seems to require a careful consideration
of the methods to use.

The particular distribution to which we presently wish to fit a linear
function has the following characteristics:

\begin{itemize}

\item The $X$ and $Y$ variables are photometric colors, obtained by
  subtracting the magnitudes of a star in two bands. Both variables
  are therefore subject to photometric errors that increase with
  magnitude but not in an entirely predictable way with color.

\item Because we observe (parts of) a non-homogeneous cloud, the
  amount of extinction is not the same for all stars, and since most
  of the area is at low extinction, the distribution of points in a
  color-color diagram is denser in the bluer end, and scarcer in the
  redder end.

\item The $X$ and $Y$ colors will usually have one band in common
  ({\it e.g.}, $J-H$ and $H-K$), which causes the errors to be
  correlated (anti-correlated in the example).

\item Because we observe the colors of a random sample of stars
  background to the cloud, the data will have intrinsic scatter caused
  by the range in spectral types of the stars. The intrinsic scatter
  is unrelated to extinction.

\end{itemize}

To summarize, the position of each datapoint in the color-color
diagram is determined by three factors: the intrinsic scatter from the
dispersion in spectral types of field stars, the reddening caused by
dust extinction, and the measurement (photometric) error. The first
alone would trace the loci occupied by the unreddened colors of the
population of stars in the field, mostly giant and main-sequence
stars; the effect of extinction is to (dim the objects and) move each
point along the line whose slope we want to determine; and the
photometric errors scatter the points in an ellipse around the
intrinsic, reddened position (not a circle because the errors are
correlated). A proper method to measure the slope of this distribution
should be robust enough to disentangle these effects and return the
single slope of the reddening vector.


\subsection{Methods for linear regression}
\label{sec:methods}

In this paper we test the following methods for linear regression
applied to the problem described in the previous section.

\subsubsection{Least squares fitting}
\label{sec:least-squar-fitt}

The ordinary least-squares (OLS) fit is the simplest approach to
linear regression. It works by minimizing the vertical distance of all
data points to successive lines in the 2-D (slope and intercept)
parameter space. Formally, it is equivalent to finding $\alpha_\mathit{OLS}$
and $\beta_\mathit{OLS}$ that minimize the quantity:

\begin{equation}
  \label{eq:3}
  \chi^2(\alpha_\mathit{OLS}, \beta_\mathit{OLS})=\sum_{i=1}^{N}(y_i - \alpha_\mathit{OLS} - \beta_\mathit{OLS} x_i)^2
\end{equation}

\noindent where $(x_i, y_i)$ is the $i^{th}$ data point, and $\alpha$
and $\beta$ are the intercept and the slope one is trying to
find. This method treats all data points the same, even though the
position of some points will be more uncertain than that of
others. The knowledge of the measurement errors for each data point
can be used to attribute different weights to the different data
points, thus optimizing the fit, which is equivalent to finding
$\alpha_\mathit{WLS}$ and $\beta_\mathit{WLS}$ that minimize the quantity:

\begin{equation}
  \label{eq:4}
  \chi^2(\alpha_\mathit{WLS}, \beta_\mathit{WLS})=\sum_{i=1}^{N}\frac{(y_i - \alpha_\mathit{WLS} - \beta_\mathit{WLS} x_i)^2}{\sigma_{yi}^2 + \beta_\mathit{WLS}^2\sigma_{xi}^2}
\end{equation}

\noindent where $\sigma_{xi}$ and $\sigma_{yi}$ are the measurement
errors of the $i^{th}$ data point. This method is called the weighed
least squares (WLS) method.


\subsubsection{Symmetrical methods}
\label{sec:symmetrical-methods}

Applications of the least squares method have been suggested to be
most appropriate when the intrinsic scatter in the data dominates over
the measurement errors (see \citet{Isobe:1990aa} for a detailed
discussion and formal description). As opposed to the method described
above, these treat both variables symmetrically, in such a way that it
is no longer meaningful to speak of dependent and independent
variables.

Two of these methods are based on calculating the OLS slope of the Y
$vs.$ X distribution, $\hat{\beta_1}=$~OLS(Y$|$X) following the
notation of \citet{Isobe:1990aa}, and that of the X $vs.$ Y
distribution, $\hat{\beta_2}=$~$1/$OLS(X$|$Y). The best fit is the
line that bisects the OLS(Y$|$X) and the OLS(X$|$Y) in the bisector
method (eq. \ref{eq:5}), or the geometric mean of the OLS(Y$|$X) and
OLS(X$|$Y) slopes in the geometric mean method (eq. \ref{eq:6}). A
third method, orthogonal regression (eq. \ref{eq:7}), minimizes the
distances of the points to a model line, but perpendicularly to the
line instead of vertically as the regular least-squares
method. \citet{Isobe:1990aa} clearly state that these three methods do
not produce the same or equivalent solutions. The plus or minus in the
equations below refer to the sign of the covariance of the two
variables | positive if they are correlated, negative if
anti-correlated.

\begin{equation}
  \label{eq:5}
  \hat{\beta}_\mathit{bisector}=\frac{\hat{\beta}_1\hat{\beta}_2-1+\sqrt{(1+\hat{\beta}_1^2)(1+\hat{\beta}_2^2)}}{\hat{\beta}_1+\hat{\beta}_2}
\end{equation}

\begin{equation}
  \label{eq:6}
  \hat{\beta}_\mathit{geom}=\pm\sqrt{\hat{\beta}_1\hat{\beta}_2}
\end{equation}

\begin{equation}
  \label{eq:7}
  \hat{\beta}_\mathit{orth}=\frac{\hat{\beta}_2-\hat{\beta}_1^{-1}\pm\sqrt{4+(\hat{\beta}_2-\hat{\beta}_1^{-1})^2}}{2}
\end{equation}

\vspace{0.4cm}



\subsubsection{Binning}
\label{sec:binning}

Binning data always causes loss of information, but, sometimes, what
is lost in information, is gained in simplicity of analysis and
results. In their studies of extinction by molecular clouds,
\citet{Lombardi06} developed a method to determine the extinction law
that consisted in binning the colors in both axes and fitting the bins
using the weighed least-squares fitting described in sect.
\ref{sec:least-squar-fitt}. The characterization of the data changes
entirely, because (1) the data points are no longer measurements but
the weighed average of many measurements, (2) the errors are no longer
measurement errors but the dispersion of colors within each bin, and
(3) the fewer high extinction points are given more weight than before
as they are represented by a bin with the same weight as those bins
containing more points. The latter point implies, although with the
potential of introducing small number statistics issues, that the
weight of any extinction range is the same, regardless of where the
majority of datapoints lie. In this way, instead of the fit being
weighed by the more abundant points at low extinction, as is the case
for the fit to all datapoints, it is equally weighed by a much larger
range of extinction thus allowing a better constrain of the extinction
law.  It also implies that part of the intrinsic scatter problem is
eliminated as the high extinction population will be dominated by
giants whose range in intrinsic colors is much narrower.

For the purpose of these experiments we have binned the data in two
ways: (1) along the color in the $X$-axis in bins of color, and (2)
iteratively along an assumed reddening vector in bins of $A_V$
following the method described by \citet{Lombardi06}.

\subsubsection{BCES method}
\label{sec:bces-methods}

The standard BCES method \citep{Akritas:1996fk} starts from the simple
observation that the slope $\beta$ that minimizes the standard OLS
Eq. (1) can be alternatively written as

\begin{equation}
\beta_\mathit{OLS} = \frac{\mathrm{Cov}(x,y)}{\mathrm{Var}(x)} \; ,
\end{equation}

\noindent where $\mathrm{Cov}(x,y)$ is the observed covariance between
the data $\{ x_i \}$ and $\{ y_i \}$, and $\mathrm{Var}(x)$ is the
variance of $\{ x_i \}$.  These two quantities can be evaluated from
the usual equations

\begin{gather}
\mathrm{Var}(x) = \frac{1}{N} \sum_{i=0}^n \bigl(x_i - \bar x \bigr)^2
\; , \\
\mathrm{Cov}(x) = \frac{1}{N} \sum_{i=0}^n \bigl(x_i - \bar x
\bigr) \bigl(y_i - \bar y \bigr) \; ,
\end{gather}

\noindent where the bar indicates the average values.  Note that for
the specific purposes of the BCES method the variance and covariance
of the data should be evaluated using $N$ in the denominator (instead
of the more common $N-1$, used when estimating the variance and the
mean from the same dataset).

The equation for $\beta$ above suggests that we can easily take into
account the presence of measurement errors on both $x$ and $y$ by
simply subtracting their effect from the estimates of the covariance
and variance.  To this purpose, suppose that each point $(x_i, y_i)$
in our dataset is affected by a statistical error, so that the
measured values $(\hat x_i, \hat y_i) = (x_i, y_i) + (\epsilon^x_i,
\epsilon^y_i)$.  The quantities $(\epsilon^x_i, \epsilon^y_i)$
represent the errors, and are drawn from some distribution; the errors
on $x$ and $y$ are not assumed to be uncorrelated here.  The presence
of the errors changes the covariance and variance in the expression of
$\beta$ above as follows

\begin{gather}
\mathrm{Cov}(x,y) \mapsto \mathrm{Cov}(x, y) +
\mathrm{Cov}(\epsilon^x, \epsilon^y) \; , \\
\mathrm{Var}(x) \mapsto
\mathrm{Var}(x) + \mathrm{Var}(\epsilon^x) \; ,
\end{gather}

\noindent where we have introduced the variance and covariance of the
errors measurements. Since the true, original variance and covariance
is relevant for the estimate of $\beta$,the equations above suggest
that in presence of measurement errors, we can replace the equation
above with

\begin{equation} \label{eq:10}
\beta_\mathit{BCES} = \frac{\mathrm{Cov}(x,y) - \mathrm(Cov)(\epsilon^x,
  \epsilon^y)}{\mathrm{Var}(x) - \mathrm(Var)(\epsilon^x)} \; .
\end{equation}

Note that, in contrast to $(x,y)$, the variance and covariance of the
measurement errors is assumed to be known and cannot be derived from
the data alone.  For example, in our case, if $(x,y)$ are two colors,
say $x = H - K$ and $y = J - H$, we will have

\begin{gather}
\mathrm{Cov}(\epsilon^x, \epsilon^y) = -\sigma^2_H \; , \\
\mathrm{Var}(\epsilon^x) = \sigma^2_H + \sigma^2_K \; .
\end{gather}

\noindent where $\sigma_H$ and $\sigma_K$ are the mean photometric
error in the $H$ and $K$ bands, respectively.

\subsubsection{A new method for linear regression: LinES}
\label{sec:new-method-lines}

The standard BCES methods is very simple to implement, but, although,
in principle, it should be able to work in presence of an intrinsic
scatter, in practice, at least in the original version presented by
\citet{Akritas:1996fk}, it is does not. The authors do mention the
possibility of accounting for intrinsic scatter in the data, but only
when $x$ is measured without error, which is not applicable to the
case of the extinction law.  Additionally, in the situations that they
consider, the amount of intrinsic scatter is taken to be unknown. In
our case, however, we can estimate this quantity directly from a
control field with no extinction: the variance and covariance of the
colors there is just the sum of the variance and covariance due to the
intrinsic scatter in color of stars, and by the photometric errors for
the control field. This way we can derive the slope using the
following expression:

\begin{equation}
\beta_\mathit{LinES} = \frac{\mathrm{Cov}(x,y) - \mathrm{Cov}(\epsilon^x, \epsilon^y) -
  \mathrm{Cov}(x^{cf},y^{cf}) + \mathrm{Cov}(\epsilon^{cfx},
  \epsilon^{cfy})}{\mathrm{Var}(x) - \mathrm{Var}(\epsilon^x) -
  \mathrm{Var}(x^{cf}) + \mathrm{Var}(\epsilon^{cfx})} \; .
\end{equation}

We call this estimate the LinES method ({\bf Lin}ear regression with
{\bf E}rrors and {\bf S}catter)\footnote{A code for fitting data with
  LinES will be available online.}. As shown in section
\ref{sec:lines-method}, this method is robust against the presence of
correlated errors in both variables, and against the presence of
intrinsic scatter.

\section{Method validation} 
\label{sec:tests}

We tested the methods described above using controlled sets of
synthetic data. The following sections discuss these data, and the
effects of varying the data parameters on the results produced by each
method.

\subsection{Synthetic data}
\label{sec:synthetic-data}

We constructed a set of simulations meant as controlled, realistic
datasets comparable to the observed data for the Pipe Nebula cores
\citep{Lombardi06,Alves07}. The simulated data consisted of a number
of points (stars) characterized by brightnesses in three bands -
arbitrarily $J$, $H$ and $K$ - affected by a value of extinction that
obeys a predetermined extinction law. We simulated three sets of data
to test the different aspects of each method.

\begin{figure}
  \resizebox{\hsize}{!}{\includegraphics{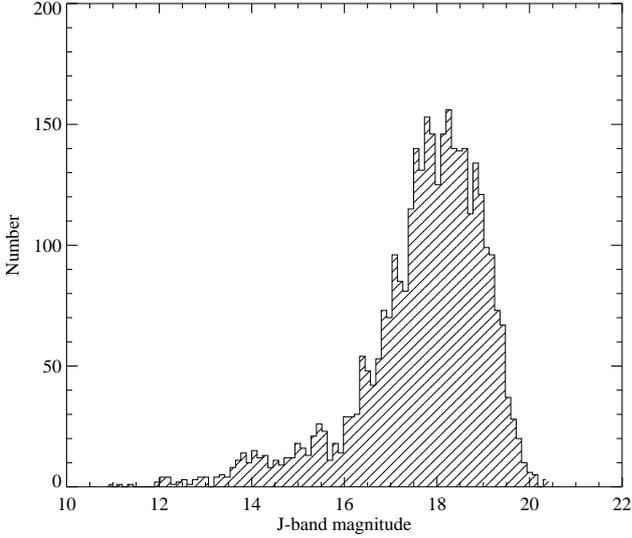}}
 \caption{Observed $J$-band luminosity function of a control field,
    used as a model distribution for the unreddened $J$ luminosities
    of the synthetic data.}
\label{fig:model-jlf}
\end{figure}

For all sets, the $J$ brightnesses were drawn randomly from the
distribution of $J$ luminosities from an observed unreddened control
field toward the galactic bulge (Fig. \ref{fig:model-jlf}).

\begin{figure}
  \resizebox{\hsize}{!}{\includegraphics{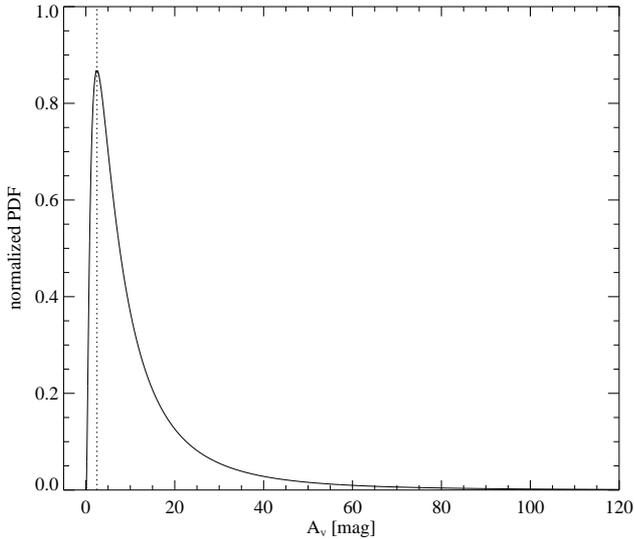}}
  \caption{Log-normal model distribution of extinction $A_V$ for a
    generic cloud. The {\it dotted line} corresponds to $A_V=2.5$ mag,
    the median extinction of the Pipe Nebula.}
\label{fig:model-av}
\end{figure}

The synthetic extinction profile was defined as a lognormal
distribution (Eq. \eqref{eq:1}) centered at $\mu=\log(2.5)$, the
logarithm of the median extinction of the Pipe, and with a width of
$\sigma=0.46$, chosen so that the number of stars at high extinctions
approximately matches that for the observed data of the Pipe cores
(Fig. \ref{fig:model-av}).

\begin{equation}
  \label{eq:1}  \mathrm{PDF} = \frac{1}{\sqrt{2\pi\sigma^2}}e^{-\frac{(\log(A_V)-\mu)^2}{2\sigma^2}}.
\end{equation}

\noindent Because the probability distribution function (PDF) for the
extinction is a lognormal, there will be many stars at low extinction
and progressively fewer stars at high extinction.

$A_K$ was then calculated from the relation $A_K/A_V = 0.112$
\citep{RiekeLebofsky85} and the values of extinction drawn from
eq. \eqref{eq:1} were applied to the $J$, $H$ and $K$ brightnesses
using the extinction law characterized by:

\begin{gather}
    \label{eq:2}
    A_H/A_K = 1.55 \\
    \label{eq:2.1}
    A_J/A_K = 1.55(\beta + 1) - \beta.
\end{gather}

\noindent The second equation, derived from eq. \eqref{eq:9} for a
$(J-H)$ {\it vs.}  $(H-K)$ diagram, defines $\beta=E(J-H)/E(H-K)$,
which means the reddening vector has a slope of $\beta$ in a $(J-H)$
{\it vs.} $(H-K)$ color-color diagram. $A_H/A_K=1.55$ is adopted from
\citet{Indebetouw05}.

\subsubsection{Set 1: Homoscedastic data, no intrinsic scatter}
\label{sec:set1}

For the first set, the $H$ and $K$ brightnesses were derived from $J$
assuming that all stars have the typical intrinsic color of giants
\citep[$J-H = 0.7$ and $H-K = 0.15$, ][]{BesselBret88}\footnote{The
  exact value of these colors is not relevant for the tests.}. Each
star was then assigned a value of extinction drawn randomly from the
extinction profile (eqs. \eqref{eq:1} - \eqref{eq:2.1},
Fig. \ref{fig:model-av}).

In addition to reddening, each star was assigned errors in $J$, $H$
and $K$, to simulate the photometric errors inherent to real data. The
errors were first applied in the simplest possible way: independently
of brightness. The magnitudes of the errors were drawn randomly and
independently for $J$, $H$ and $K$, from a normal distribution with
$\mu = 0$ and $\sigma = 0.05$, so that 95\% of the stars will have
errors below 0.1 mag, the typical acceptable errors for photometry. A
random value from these distributions was then added to the magnitudes
of each star in all bands independently. The synthetic data produced
in this way will henceforth be referred to as Set 1, or homoscedastic
dataset, since the errors do not scale with the variables.

\subsubsection{Set 2: Heteroscedastic data, no intrinsic scatter}
\label{sec:set2}

\begin{figure}
  \resizebox{\hsize}{!}{\includegraphics{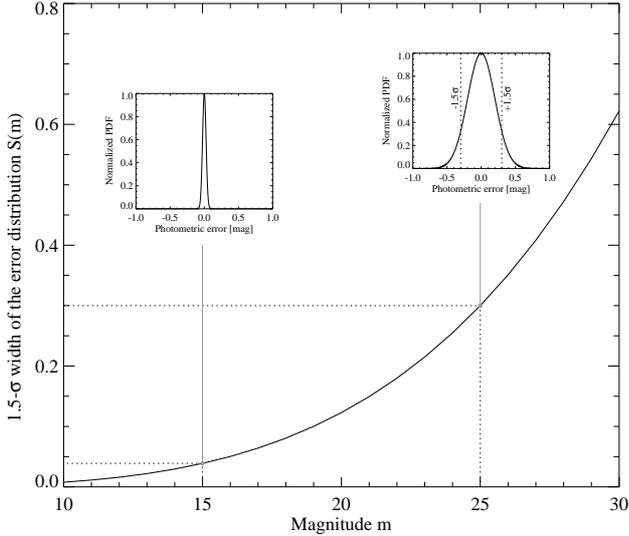}}
  \caption{Width of the error distribution as a function of magnitude
    used to generate the ``photometric errors''. The insets show the
    actual error distribution for stars of magnitude 15 ({\it left})
    and 25 ({\it right}).}
\label{fig:model-err}
\end{figure}

A second approach was designed to reproduce the fact that the error in
real observations does in fact increase with magnitude. We modeled
this dependence using an error distribution in the form of a power-law
$S(m) = Cm^x$, where $S$ is the typical error associated with
magnitude $m$ (Fig. \ref{fig:model-err}). The normalization constant
$C$ was set so that 90\% of the $25^{th}$ magnitude stars have an
error up to 0.3 mag, and the index $x$, that defines how rapidly the
errors increase with magnitude, was set to $4$. Both parameters were
empirically chosen to produce a curve that resembles a typical error
distribution of the NIR data, but other combinations around these
values would also be good representations of the general distribution
of photometric errors in a sample, and do not change the results. This
function was then used to determine the width of the Gaussian from
which the errors for each star were drawn according to its magnitude,
the error for the bright stars being drawn from a narrow Gaussian, and
that for the faint stars being drawn from a wider Gaussian (see the
insets in Fig. \ref{fig:model-err}). The synthetic data produced in
this way will hereafter be referred to as Set 2, or heteroscedastic
data, since the errors do scale with magnitude.

\begin{figure*}
\includegraphics[width=\textwidth]{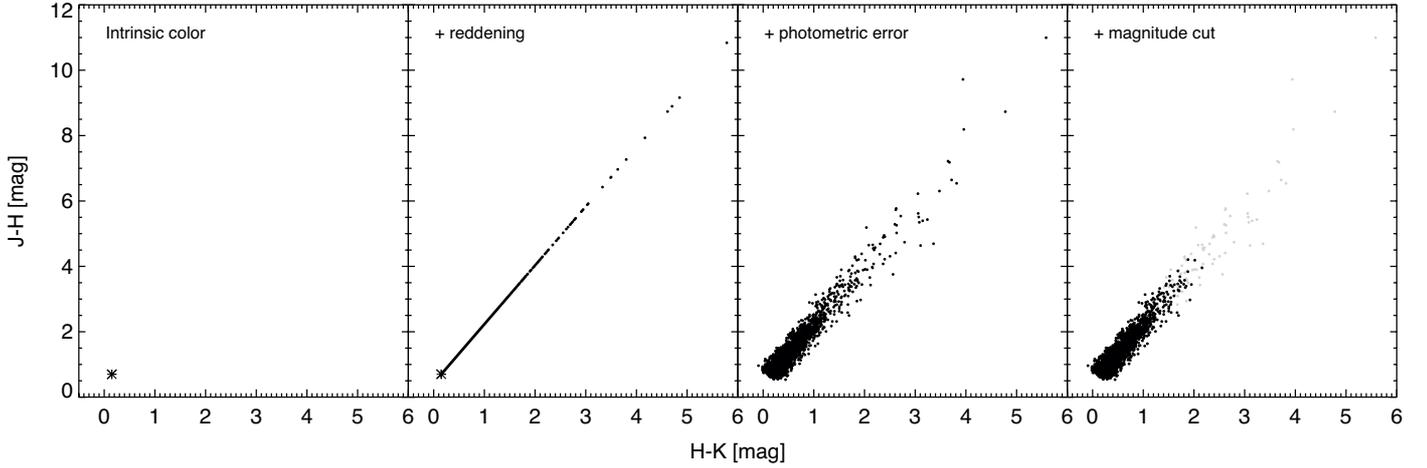}
\caption{Example of the generation of synthetic heteroscedastic data
  without scatter. From left to right: all synthetic stars have the
  same intrinsic color; they are reddened according to the log-normal
  distribution of $A_V$; photometric error that depends on magnitude
  is added; and finally a magnitude cut is implemented, reducing the
  size of the sample and the coverage in $A_V$.}
\label{fig:gen-synth-data}
\end{figure*}

Figure \ref{fig:gen-synth-data} shows the construction of the
synthetic data in the color-color diagrams for a given realization of
heteroscedastic data without intrinsic scatter.

\subsubsection{Set 3: Heteroscedastic data with intrinsic scatter}
\label{sec:set3}

The final experiment was meant to represent the fact that the stars
being observed through the cloud do not have a single color, but are
rather distributed along the main-sequence and giant stars loci
according to their individual masses and ages. This intrinsic
``scatter'' about a single colour was modeled by assuming a fraction
$f$ of main-sequence-to-giant stars, and randomly populating the loci
accordingly, regardless of the stars' brightnesses. This does not
produce a completely realistic set of data because the position of the
stars would depend on their brightness (see
Sect. \ref{sec:real-data}), but our goal is to test how the presence
of a generic, non-symmetrical intrinsic scatter affects the
measurement of the slope of the extinction law.

In this dataset there are then three contributions to the distribution
of points in the color-color diagram: the distribution of intrinsic
colors of stars background to the cloud, that spread the stars along
the main-sequence and giant loci; the extinction, that moves each
point from its intrinsic position along a line whose slope is
determined by the properties of the dust; and the magnitude-dependent
photometric error, that, in a color-color diagram, is equivalent to
each point being drawn from an ellipse around each intrinsic and
reddened point, whose dimensions depend on the observed brightness of
the corresponding star.

The data produced in this way will be referred to Set 3, or
heteroscedastic data with intrinsic scatter.

\subsubsection{Synthetic control fields}
\label{sec:synth-control-fields}

For each set intended to pose as science data, we generated a
corresponding set intended to pose as data from a control field. For
Sets 1 and 2 the synthetic control fields had equivalent homoscedastic
or heteroscedastic errors, respectively, drawn randomly but
independently from the same distributions as the errors for the
synthetic ``science'' fields. For Set 3, apart from the
heteroscedastic errors, the control field simulation also included an
amount of intrinsic scatter equivalent to that of the synthetic
``science'' field. Neither of the synthetic control fields included
extinction, as they are meant to represent the population background
to the cloud that is causing the reddening on the ``science'' field
stars.

\subsection{``Real'' data}
\label{sec:real-data}

There is one aspect of real observations that we cannot test with the
synthetic datasets described above: whereas at low extinction both
main-sequence and giants can be observed, at high extinction the
main-sequence stars are more efficiently dimmed below the detection
limit, leaving a population dominated by the intrinsically brighter
giants at redder colors. In our definitions for the synthetic data,
this means that $f$ should change with extinction, being larger at low
extinctions and progressively smaller at high extinctions. As already
hinted in Sect. \ref{sec:set3}, in set 3 of our synthetic data we do
simulate intrinsic scatter, but the brightness of each star does not
scale with its spectral type, which translates into having a constant
$f$ throughout the entire extinction range. To test the effect of a
varying amount of intrinsic scatter with extinction on the methods we
applied them to actual observations of control fields (courtesy of
C. R\'oman-Z\'u\~niga), which we reddened in the same way as we did
the synthetic data. We refer to these datasets as ``real'', keeping
the quotes to make clear that the actual observed data were then
artificially modified.

The first dataset contains data taken with the SOFI instrument at
ESO's New Technology Telescope, in the direction of the galactic disk
($10^{h}38^{m}12^{s}$, $-59^\circ12'02''$, J2000.0), on the night of
March 31$^{st}$, 2006. This dataset, which we refer to as ``disk
dataset'', contains 548 stars with PSF photometry in the $J$, $H$ and
$K_s$ filters. This is not the ideal dataset for two reasons: first,
it contains few stars, and second, since this is a field in the
galactic disk, it already has some extinction.

The second dataset contains data also taken with the SOFI instrument
at ESO's New Technology Telescope on the night of June 22$^{nd}$,
2002, but in the direction of the galactic bulge
($17^{h}08^{m}10^{s}$, $-28^\circ03'03''$, J2000.0). This dataset,
which we refer to as ``bulge dataset'', contains 1071 stars with PSF
photometry in the $J$, $H$ and $K_s$ filters.
  
5000 ``science'' subsets were drawn randomly from each of these
datasets, and extinction was applied to each star from the extinction
distribution (see Eq. \eqref{eq:9}, Figure
\ref{fig:model-av}). Similarly, 5000 ``control field'' subsets were
drawn randomly from each dataset, and used as they were. 450 and 800
stars were drawn from the disk and the bulge datasets, respectively,
in order to keep the possibility of the ``control'' and ``science''
subsets being made of different stars.

\begin{figure*}
  \includegraphics[width=\textwidth]{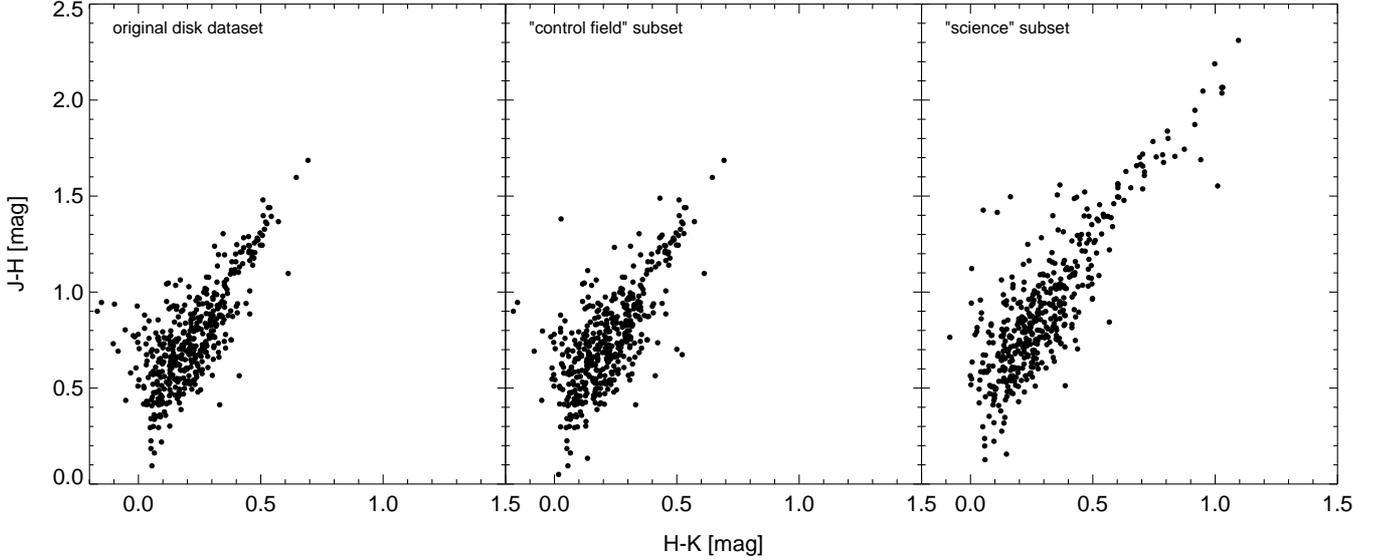}
  \caption{$(J-H)$ {\it vs.} $(H-K)$ color-color diagrams for the
    disk dataset. {\it Left:} all stars in the dataset. {\it
      Middle:} 450 stars chosen randomly from the dataset,
    representing the ``control field''. {\it Right:} 450 stars
    chosen randomly from the dataset, and reddened to represent the
    ``science field''.}
  \label{fig:disk-dataset}
\end{figure*}

\begin{figure*}
  \includegraphics[width=\textwidth]{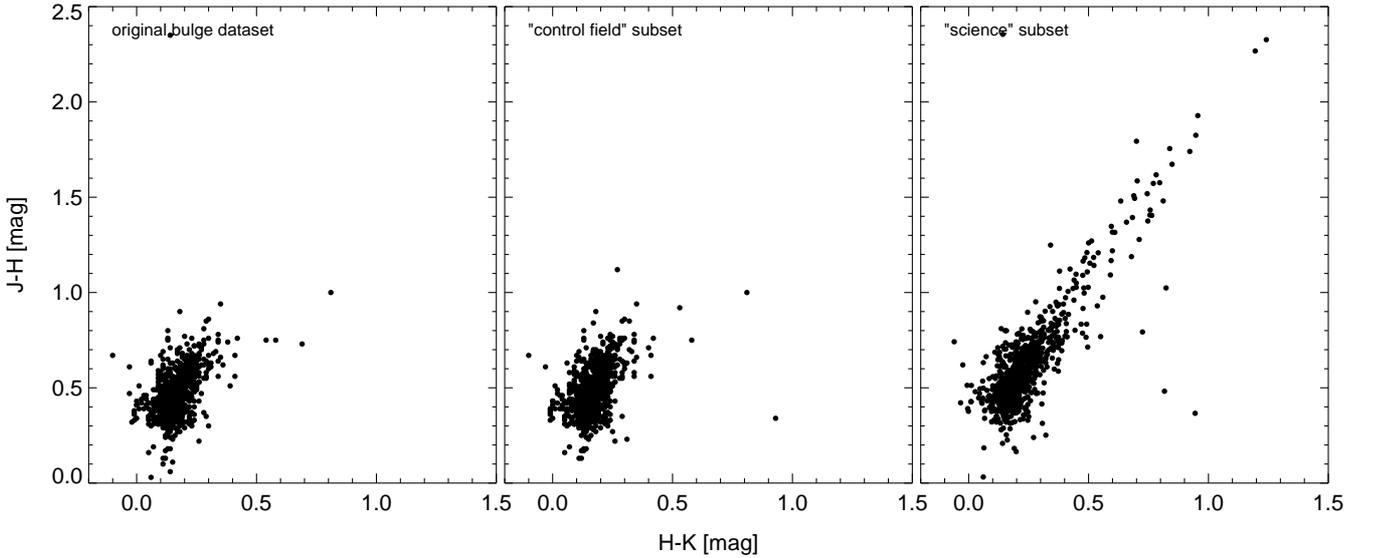}
  \caption{$(J-H)$ {\it vs.} $(H-K)$ color-color diagrams for the
    bulge dataset. {\it Left:} all stars in the dataset. {\it
      Middle:} 800 stars chosen randomly from the dataset,
    representing the ``control field''. {\it Right:} 800 stars
    chosen randomly from the dataset, and reddened to represent the
    ``science field''.}
  \label{fig:bulge-dataset}
\end{figure*}

For the effects of these experiments, the two sets differ in three
aspects: the bulge dataset contains more stars, does not have
extinction, and is made up mostly of giant stars, whereas the disk
dataset contains fewer stars, has already some extinction, and is more
likely to have a higher fraction of main-sequence objects.  The
$(J-H)$ {\it vs.} $(H-K)$ color-color diagrams of the two datasets are
shown in Figures \ref{fig:disk-dataset} and \ref{fig:bulge-dataset},
and illustrates these differences. In both figures, the left panels
are the original datasets, and the middle and right panels show one
realization of the extracted ``control field'' and ``science'' subsets
used for the tests, respectively.

The only caveat regarding these datasets is that, when reddening the
stars to pose as science subsets, their magnitudes change, so their
associated errors should change accordingly. However, since these are
real observations and given the statistical nature of the errors, that
adjustment is not possible. In practice, this means that there will be
stars at high extinction with an underestimated associated error, but
for the effect of our tests this is not critical, since there
continues to be no clear dependence of the error with extinction, as
is the case in real data subject to extinction.

\subsection{Parameters}
\label{sec:synth-parameters}

To test the robustness of the methods, the synthetic datasets were
generated using a range of parameters, namely input slopes of the
reddening vector, amount of intrinsic scatter, size of the sample, and
other specific parameters only relevant to some of the methods.

\paragraph{Input slope}
\label{sec:input-slope}
Each set was generated with seven values of input slope $\beta$ in the
range $[-1.0, 3.0]$ to cover the range expected for an extinction law
in the near- and mid-infrared. The methods were tested under ideal
conditions of number of stars and $A_V$ coverage to test only the
ability of the methods to deal with different values of $\beta$. The
input slope was varied to guarantee that our conclusions are not only
valid for one specific value of $\beta$. While varying the remaining
parameters the input slope was fixed at 1.8.

\paragraph{Magnitude limit}
\label{sec:magnitude-limit}
The magnitude limit was parameterized by $m_c$. It corresponds to
setting a brightness limit in real data, and it was applied
identically in $J$, $H$ and $K$ such that all the synthetic stars
fainter than $m_c$ in any band after applying the extinction, are
discarded from the fit (see rightmost panel in
Fig. \ref{fig:gen-synth-data}). A magnitude limit is naturally set in
real data (detection limit), but is also something one might consider
doing artificially to eliminate those stars with the largest
photometric errors.

Decreasing $m_c$ is equivalent to reducing the size of the sample,
both in number and in range of extinction (the stars at high
extinction will likely be fainter), while simultaneously reducing the
range of errors in Sets 2 and 3 (fainter stars will have larger
errors). By construction, a value of $m_c=25$ corresponds to allowing
the largest errors to be 0.3 mag (see sect. \ref{sec:set2}).

\begin{table}
  \caption{Effective number of stars as a function of magnitude cut.} 
  \label{tab:nstars_eff}
  \centering
  \begin{tabular}{c c c c c c c c c}
   \hline \hline
    $m_c$ & & $N_S$ & & $N_S^{\mathit{eff}}$  & & $r_{\mathit{sci}}$ &
    {\it{(H$-$K)}$_{\mathit{max}}$}  \\
    \hline
    25 & & 5000 & & 4929 & & 0.99 & 2.32 \\
    23 & & 5000 & & 4850 & & 0.97 & 1.98 \\
    21 & & 5000 & & 4550 & & 0.91 & 1.65 \\
    19 & & 5000 & & 2930 & & 0.59 & 1.35 \\
    17 & & 5000 & & 680 & & 0.14 & 1.04 \\
    \hline
  \end{tabular}
\end{table}

Table \ref{tab:nstars_eff} shows the average\footnote{Average obtained
  from 5000 realizations.} number of stars from Set 3 that survive
each magnitude cut (effective number of stars, $N_S^{\mathit{eff}}$)
from an initial sample of $N_S = 5000$ synthetic stars. The ratio
$r_{\mathit{sci}}=N_S^{\mathit{eff}}/N_S$ reflects the functional form
of the model distributions and, as such, would be the same for any
other input number of stars $N_S$ for each magnitude cut. The table
also shows the average $H-K$ color of the most heavily reddened
datapoint attained for each magnitude cut, illustrating the loss in
$A_V$ coverage with magnitude limitation. The same magnitude cuts were
applied to the control field, but because the control field does not
have extinction, the effect of the cut in the effective number of
stars is much more subtle. The ratio of effective to initial number of
stars for the control field is $1.00$ for $m_c \le 21$ mag, and drops
to $0.90$ and $0.25$ for $m_c$ of 19 mag and 17 mag, respectively.

\paragraph{Number of stars}
\label{sec:number-stars}
Generating fewer stars in the first place also changes the size of the
sample. The (subtle) difference with respect to implementing magnitude
cuts is that the sample with fewer stars and no magnitude cut will
most likely have a broader range of $A_V$ than would a richer sample
with magnitude cut, as some of the fewer stars that are generated and
kept may still be faint and heavily extincted from the random drawing
process. In real observations, generating fewer stars without imposing
magnitude cuts would be comparable to observing a sparse region of the
sky where the faint and highly reddened stars can be
detected. Imposing a magnitude limit, on the other hand, would
correspond to having shallow observations regardless of the richness
of the observed field; in this case, the fainter and more reddened
stars would not be detected.

We tested the methods against varying number of stars in the range
$[100, 5000]$, both in the science and the control field.

\paragraph{Amount of intrinsic scatter}
\label{sec:amount-intr-scatt}
The amount of intrinsic scatter for Set 3 was parametrized by $f$, the
fraction of stars in the main-sequence locus with respect to those in
the giant locus, taking values of 0.01, 0.15 and 0.50. Given the shape
of the loci, a larger scatter is obtained for larger values of $f$,
although the giant locus alone still produces some intrinsic
scatter. Larger values of $f$ would correspond to a large fraction of
the stars behind the cloud being main-sequence in real data, as
opposed to having mainly giants. When varying the remaining
parameters, $f$ is fixed at 0.15 for Set 3 ($f$ is not a parameter in
Sets 1 and 2).

\paragraph{Number of control-field stars}
\label{sec:number-control-field}
The synthetic control field is only used for the LinES method, as the
other methods rely solely on the science data. Unless explicitly
stated, the control field was generated with the same number of stars
and was subject to the same magnitude cuts as the synthetic
``science'' datasets. This means that, after applying a given
magnitude cut, there will effectively be more stars in the control
field sample than in the science sample, since a fraction of the stars
in the science dataset will have been dimmed below the magnitude cut
by the effect of extinction. This mimics real observations in that the
control field should be obtained from a region of comparable
background stellar density (same number of stars generated in the
simulations) and using the same instrumental setup (same magnitude cut
in the simulations) as the science field. To test the effect of a less
than ideal control field, we tested the LinES method against varying
numbers of stars also in the control-field datasets.

\subsection{Results from synthetic data}
\label{sec:synth-results}


Each method was applied to 5000 realizations of the synthetic data,
producing an average value $\hat{\beta}$, and a dispersion
$\sigma_{\hat{\beta}}$ around the average for each parameter within
sets 1, 2, and 3. We define bias $b$ as the difference between
$\hat{\beta}$ and the input value of the slope
($b=\hat{\beta}-\beta$), and consider a method unbiased if it produces
an estimate within the 1-$\sigma$ dispersion of the input value ({\it
  i.e.}, $b \le \sigma_{\hat{\beta}}$).

We note that the quoted absolute values of the biases are formally
only valid for the conditions of these simulations, namely for the
magnitude of the error and the functional form of the error
distribution with magnitude.



The results are summarized below. Figures \ref{fig:results-ols}
through \ref{fig:results-cbces} show the bias as a function of varied
parameters.




\subsubsection{Ordinary least-squares}
\label{sec:ols}

\begin{figure*}
\centering
\includegraphics[width=17cm]{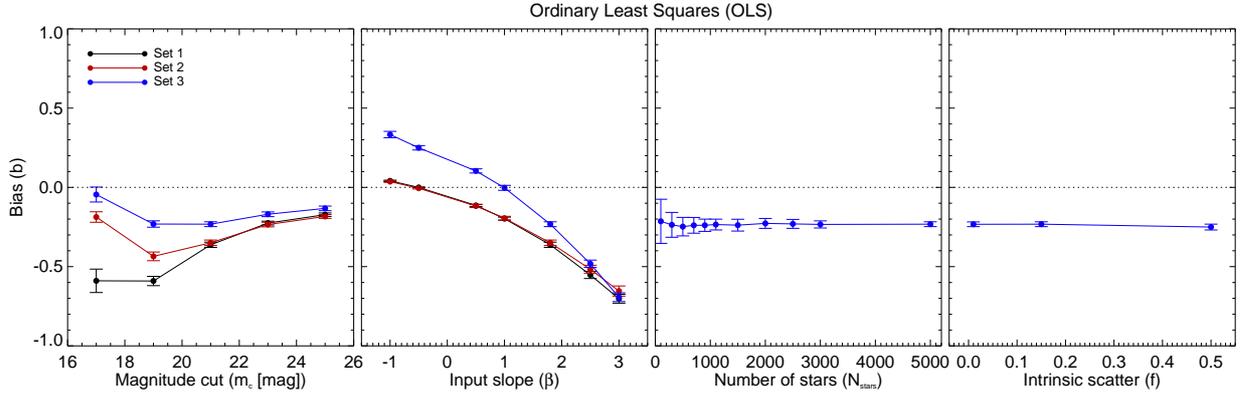}
\caption{From left to right, bias in the slope as a function of the
  magnitude cut, input slope, number of stars in the science field,
  and intrinsic scatter for the ordinary least squares (OLS)
  method. The {\it black}, {\it red} and {\it blue} lines represent
  the results for the 5000 realizations of Sets 1, 2 and 3,
  respectively. The error bars correspond to the 1-$\sigma$ dispersion
  in the results.}
\label{fig:results-ols}
\end{figure*}

The OLS method has long been known to be biased when there are errors
in both coordinates. This was observed also in our simulations
(Fig. \ref{fig:results-ols}), with the method failing to recover the
right value of the slope in the majority of the tests performed. It
did provide an unbiased result for the following combinations of input
slopes and datasets: $\beta=-0.5$ for Set 1, $\beta=-1.0$ and
$\beta=-0.5$ for Set 2, and $\beta=1.0$ for Set 3, making this method
unsuitable if one is trying to find precisely $\beta$.

This method is also not robust against variations in $m_c$, the bias
and the dispersion both increasing for brighter $m_c$ in the
heteroscedastic sample, except for the brightest magnitude cut
considered, where the bias is suddenly reduced. It also reacts,
although to a lesser extent, to changes in the amount of intrinsic
scatter in the heteroscedastic sample (Set 3), the bias increasing
with $f$.

This method is extremely robust against variations in the number of
stars within the same magnitude cut, although the dispersion in the
slope increases for fewer stars as would be expected from poor
statistics, and the absolute value is biased.

Overall, this method is not a reliable estimator of the extinction
law.

\subsubsection{Weighed least-squares}
\label{sec:wols}

\begin{figure*}
  \centering
\includegraphics[width=17cm]{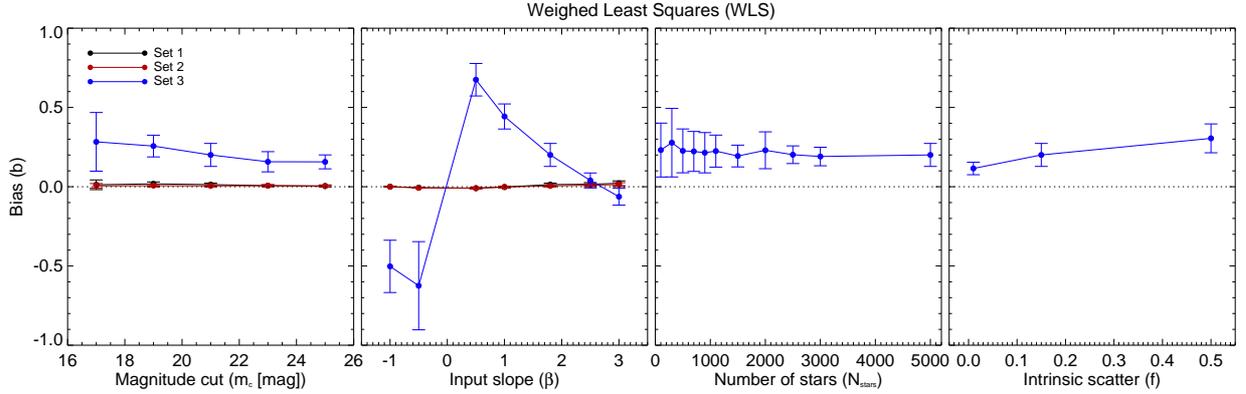}
\caption{From left to right, bias in the slope as a function of the
  magnitude cut, input slope, number of stars in the science field,
  and intrinsic scatter for the weighed least squares (WLS)
  method. The {\it black}, {\it red} and {\it blue} lines represent
  the results for the 5000 realizations of Sets 1, 2 and 3,
  respectively. The error bars correspond to the 1-$\sigma$ dispersion
  in the results.}
\label{fig:results-wls}
\end{figure*}

The WLS method performs well in homoscedastic or heteroscedastic data
without intrinsic scatter (Fig. \ref{fig:results-wls}), suggesting
that, under these conditions, the method can deal properly with the
presence of errors in both coordinates, and with them being
correlated. For these samples the dispersion in the slope is
remarkably small, making it a very accurate estimator. In the presence
of intrinsic scatter, however, the method systematically fails to
recover the input slope whatever its value in the range $[-1.0, 3.0]$,
although it does come close around $\beta=2.5$. The bias as a function
of input slope plot for this method and dataset
(Fig. \ref{fig:results-wls}, blue line) suggests that there could be
another unbiased value of $\beta$ between $-0.5$ and $0.5$, but tests
suggest that there is instead a discontinuity around $\beta=0$.

Although biased for Set 3, this method is robust against variations in
the number of stars to within 1.5\% in the considered range, but the
dispersion increases steadily for fewer stars.

Biased in the presence of even a small intrinsic scatter, this method
is not a reliable estimator of the extinction law.

\subsubsection{Symmetrical methods}
\label{sec:symm-methods-results}

\begin{figure*}
\centering
\includegraphics[width=17cm]{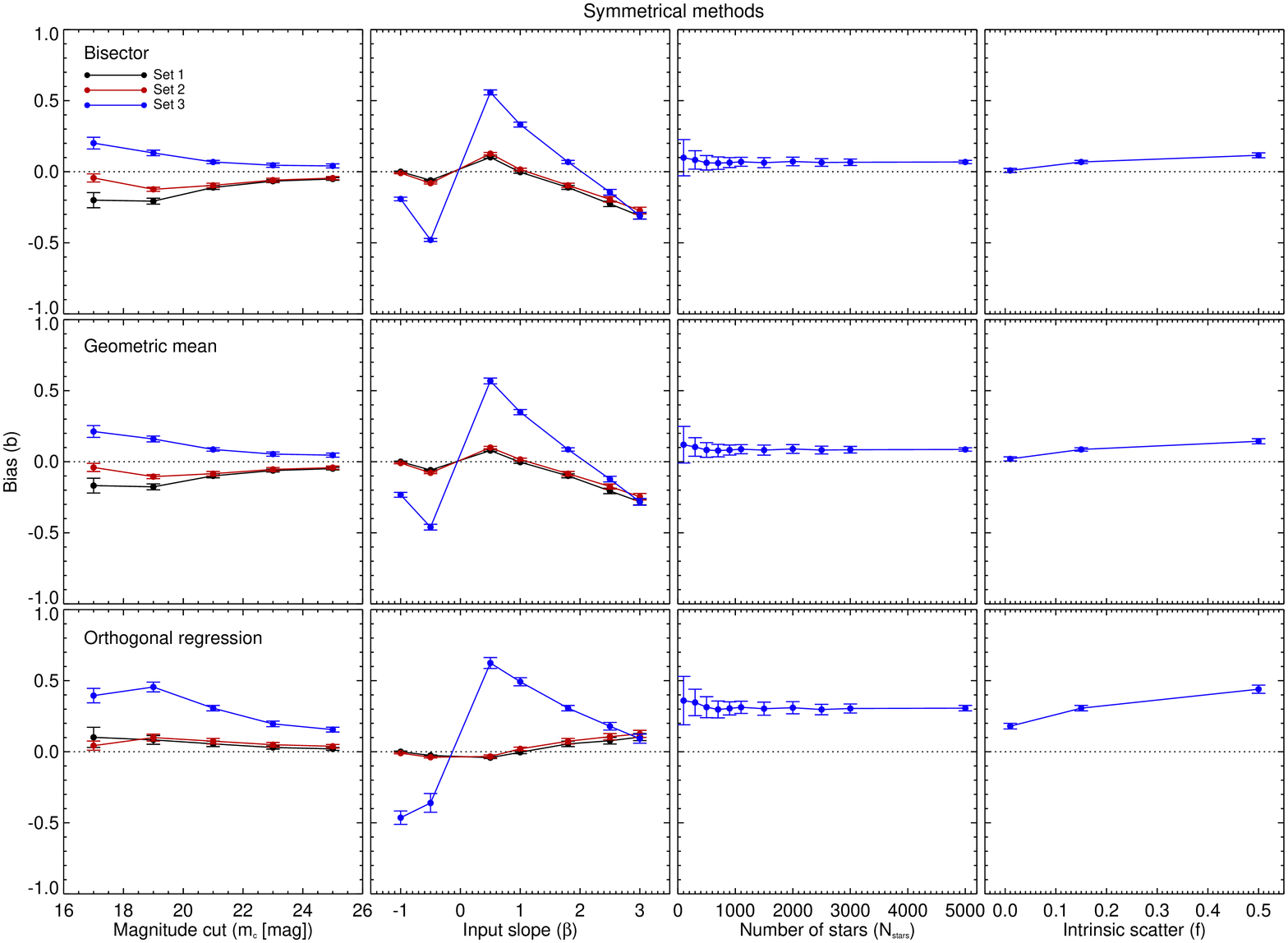}
\caption{From left to right, bias in the slope as a function of the
  magnitude cut, input slope, number of stars in the science field,
  and intrinsic scatter for the symmetrical methods: the bisector
  method in the top four panels, the geometric mean method in the
  middle panels, and the orthogonal regression method in the bottom
  panels. The {\it black}, {\it red} and {\it blue} lines represent
  the results for the 5000 realizations of Sets 1, 2 and 3,
  respectively. The error bars correspond to the 1-$\sigma$ dispersion
  in the results.}
\label{fig:results-symm}
\end{figure*}

The three symmetrical methods returned biased results for all tests
(Fig. \ref{fig:results-symm}), except for input slopes of $-1.0$ and
$1.0$ in datasets 1 and 2 (without intrinsic scatter). The bisector
and geometric mean methods produce very similar results. The
orthogonal regression method presents the largest biases of the three.

Besides being biased, neither method is robust against variations in
$m_c$ or $f$ in the presence of intrinsic scatter, the bias increasing
for bright magnitude cuts and more intrinsic scatter. The methods are
highly robust to variations in the number of stars, the dispersion
increasing steadily for fewer stars.

These methods are not reliable estimators of the extinction law.

\subsubsection{Binning in $(H-K)$}
\label{sec:binning-hk}

\begin{figure*}
\centering
\includegraphics[width=17cm]{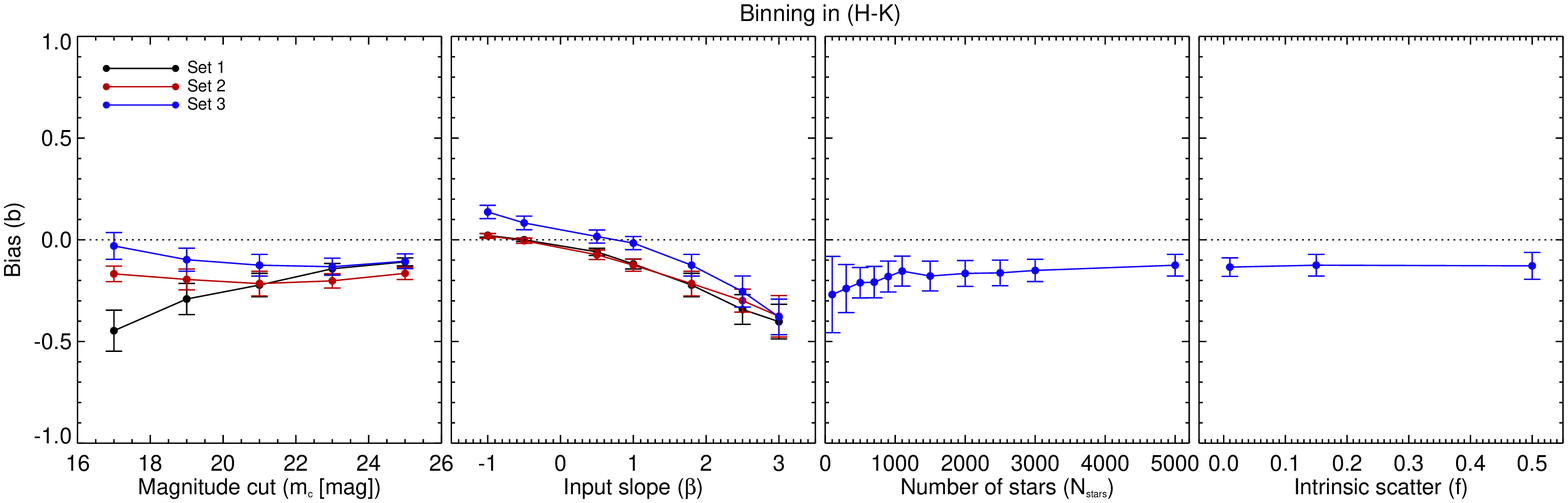}
\caption{From left to right, bias in the slope as a function of the
  magnitude cut, input slope, number of stars in the science field,
  and intrinsic scatter for the binning in $(H-K)$ method. The {\it
    black}, {\it red} and {\it blue} lines represent the results for
  the 5000 realizations of Sets 1, 2 and 3, respectively. The error
  bars correspond to the 1-$\sigma$ dispersion in the results.}
\label{fig:results-binhk}
\end{figure*}

This method is biased for most datasets and parameters tested; the
exceptions are for Set 3 with the brightest magnitude cut ($m_c=17$
mag) or for $\beta=0.5$ and $1.0$, and for $\beta=-0.5$ and no
intrinsic scatter (Fig. \ref{fig:results-binhk}).

The slope in mostly underestimated for all datasets, with the bias
increasing for brighter $m_c$ for Set 1, and keeping relatively stable
against varying $m_c$ for Sets 2 and 3. For Set 3 the bias slightly
increases with $f$, as does the dispersion. This method reacts to the
number of stars for a given magnitude cut, the bias increasing toward
fewer stars. Overall, this is not a reliable estimator of the
extinction law.

\subsubsection{Binning in $A_v$}
\label{sec:binning-a_v}

\begin{figure*}
\centering
\includegraphics[width=17cm]{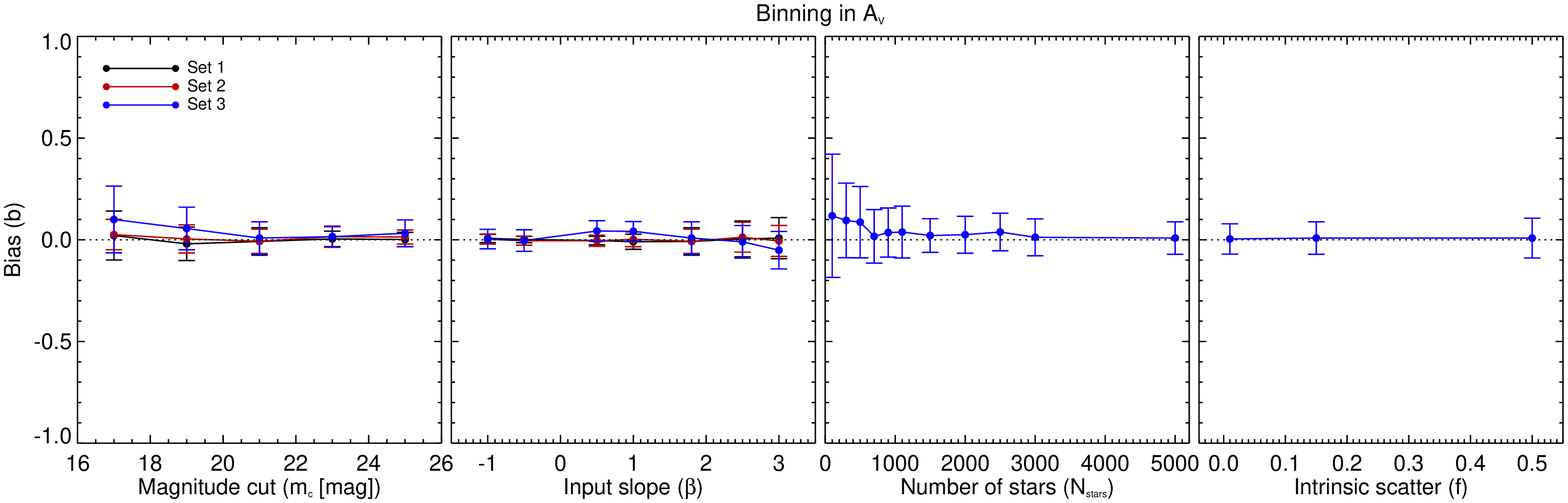}
\caption{From left to right, bias in the slope as a function of the
  magnitude cut, input slope, number of stars in the science field,
  and intrinsic scatter for the binning in $A_V$ method. The {\it
    black}, {\it red} and {\it blue} lines represent the results for
  the 5000 realizations of Sets 1, 2 and 3, respectively. The error
  bars correspond to the 1-$\sigma$ dispersion in the results.}
\label{fig:results-binav}
\end{figure*}

This method is formally unbiased for all tested values of $\beta$ when
applied to the three datasets (Fig. \ref{fig:results-binav}). However,
whereas the bias is always close to zero for the datasets without
intrinsic scatter, it becomes slightly large for most values of
$\beta$ when intrinsic scatter is introduced.

The method is robust against variations in the amount of intrinsic
scatter within the considered range, and it is stable against
variations in magnitude cut, except for the brightest value of $m_c$
for Set 3, where the smaller number of stars and range in $A_V$ result
in very few bins for the fit. The dispersion increases toward brighter
magnitude cuts and amount of intrinsic scatter. Because it is based on
binning, this method reacts significantly to the number of stars
within the same magnitude cut, the bias and the dispersion increasing
for fewer stars.

Small variations occur when the size of the bin is varied, with the
slope being increasingly overestimated for smaller bins, but the
method continues to be unbiased within the dispersion.

This method is a reliable estimator of the extinction law.

\subsubsection{BCES method}
\label{sec:bces-method}

\begin{figure*}
\centering
\includegraphics[width=17cm]{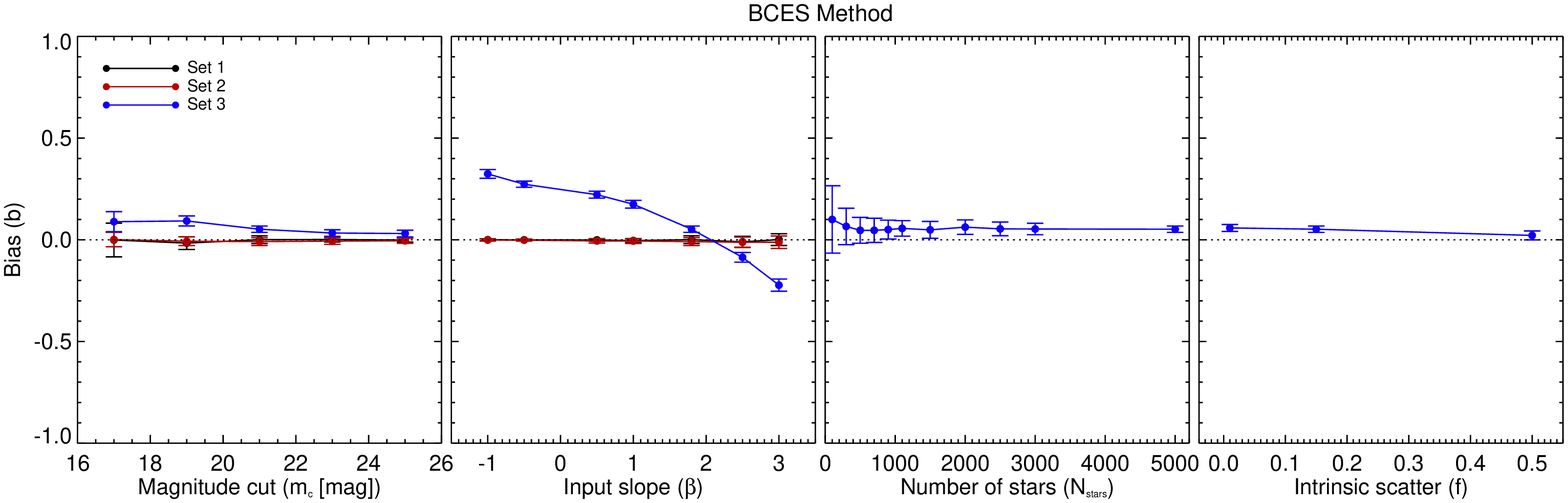}
\caption{From left to right, bias in the slope as a function of the
  magnitude cut, input slope, number of stars in the science field,
  and intrinsic scatter for the BCES method. The {\it black}, {\it
    red} and {\it blue} lines represent the results for the 5000
  realizations of Sets 1, 2 and 3, respectively. The error bars
  correspond to the 1-$\sigma$ dispersion in the results.}
\label{fig:results-bces}
\end{figure*}

This method is highly reliable and unbiased for homoscedastic data and
for heteroscedastic data without intrinsic scatter (Sets 1 and
2). However, in the presence of intrinsic scatter (Set 3), it becomes
biased for all input slope values except for $\beta=2.1$
(Fig. \ref{fig:results-bces}) for $f=0.15$. Since the other parameters
were tested using an input slope of 1.8 (very close to 2.1), the bias
seems small in the $m_c$, $N_\mathit{stars}$ and $f$ plots, but we can
nevertheless analyze the sensitivity of the method to these
parameters. For Set 3, the bias is mildly sensitive to $m_c$, robust
against variations in the number of stars, and only minimally
sensitive to variations in the amount of intrinsic scatter $f$.

Since real data will be similar, in essence, to Set 3, we do not
consider this method a reliable estimator in the specific case of the
extinction law.

\subsubsection{LinES}
\label{sec:lines-method}

\begin{figure*}
\centering
\includegraphics[width=17cm]{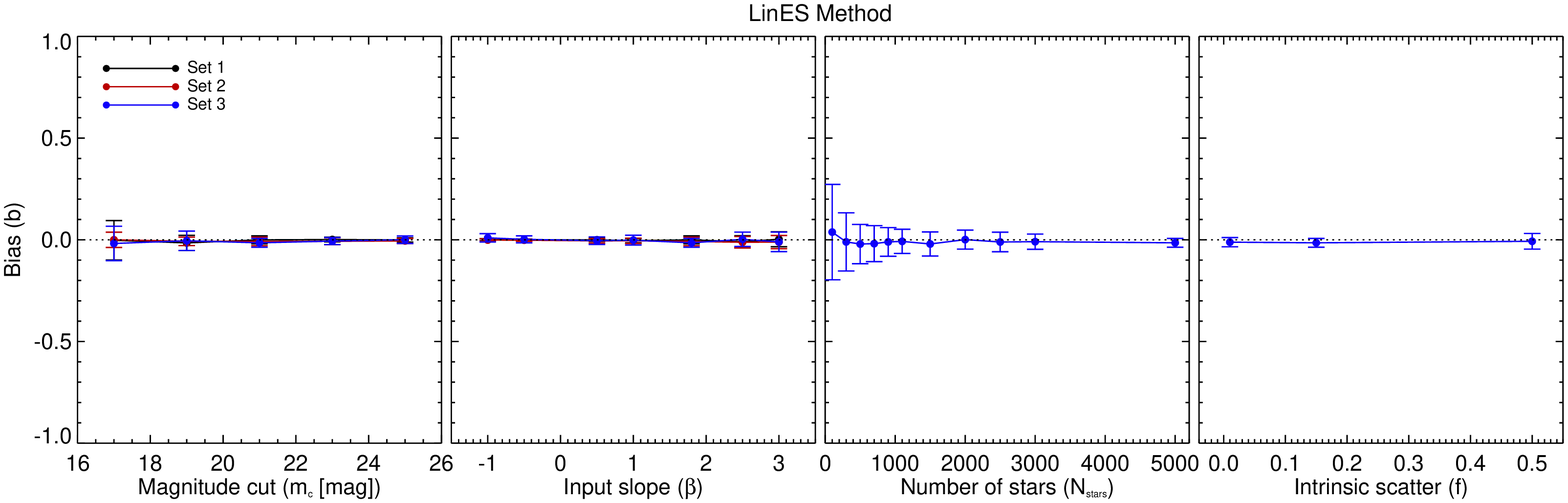}
\caption{From left to right, bias in the slope as a function of the
  magnitude cut, input slope, number of stars in the science field,
  and intrinsic scatter for the LinES method. The {\it black}, {\it
    red} and {\it blue} lines represent the results for the 5000
  realizations of Sets 1, 2 and 3, respectively. The error bars
  correspond to the 1-$\sigma$ dispersion in the results.}
\label{fig:results-cbces}
\end{figure*}

This method is the most unbiased and robust of all presented here for
homoscedastic or heteroscedastic data, with or without intrinsic
scatter, for all tested values of the input slope
(Fig. \ref{fig:results-cbces}).

The method is robust against variations in $m_c$ and $f$, number of
stars in the science field, and number of stars in the control field,
although the dispersion follows the same tendency as before: increases
for brighter magnitude cuts, slightly increases with $f$, and
decreases with number of science and/or control field stars. The bias
is larger when there are simultaneously very few science and control
field stars, and small $A_V$ coverage, but is nevertheless better than
any of the other methods for the same conditions. The dispersion is
significantly smaller than the next least unbiased method, the binning
in $A_V$, in all cases, granting it more precision.

Since LinES relies on the characterization of the data through the
properties of a control field, we tested the stability of the method
against variations in the number of stars in the control field.  Given
a reasonable number of stars in the science field, the method is
robust against variations in the number of control field
stars. However, if the science field itself does not have enough stars
or $A_V$ coverage, the bias increases further for few control field
stars. Invariably, the dispersion increases toward fewer control field
stars. This method has proven to be robust as long as the control
field is a good representation of the underlying population on the
science field, even if containing a smaller number of stars.

The excellent performance of this method while varying all relevant
parameters validates the LinES method for our case study. In general,
it will provide accurate results for observations of cores with either
rich or poor background populations, regardless of their spread in
spectral types, even for relatively shallow observations, as long as
there is a reasonable spread in extinction and the control field is a
good representation of the reddened, background population.

\paragraph{Limitations}
\label{sec:limitations}

The simulations show that LinES is not reliable for distributions that
do not cover a wide enough range of extinction, if the errors are too
large. For reasonable errors, like those described for the
simulations, LinES starts overestimating the slope by more than 10\%
for ranges in $x$-color (i.e., the color plotted on the $x$-axis)
smaller than 0.25 to 0.45 mag for slopes between 0.6 and 3.0,
respectively, and underestimating the slope by more than 10\% for the
same ranges in $x$-color for slopes between 0.3 and 0.5. This method
should therefore not be applied to data that span less than these
values in $x$-color.

\paragraph{Error estimation}
\label{sec:error-estimation}

We used the bootstrap method \citep[e.g., ][]{Wall:2003uq} to estimate
the uncertainty in the slope derived by LinES. This method consists in
randomly dividing each sample in two equal-number subsets, and
measuring the slope of the extinction law in each subset. This
produces two values of $\beta$ from which we derive the standard
deviation $\sigma_{\beta_i}$. This was repeated $N=1000$ times and the
uncertainty in the slope was defined as:

\begin{equation}
  \label{eq:8}
  \sigma_\beta=\frac{1}{\sqrt{2}N}\displaystyle\sum_{i=1}^N \sigma_{\beta_i}
\end{equation}

We applied this method on the synthetic data and compared the results
with the dispersion $\sigma_{\hat{\beta}}$ from the simulations. We
found that the bootstrap uncertainty is typically $80\%$ of
$\sigma_{\hat{\beta}}$, the dispersion from the synthetic data, and
the value we believe is a better estimate of the actual dispersion
expected for real data. Given the consistency of the results against
the variation of the different parameters, we take the uncertainty on
the LinES result of our case study to be $1.25$ times the uncertainty
derived by the bootstrap method described above.

\subsection{Results from ``real'' data}
\label{sec:real-results}

\begin{figure*}
  \centering
  \includegraphics[width=17cm]{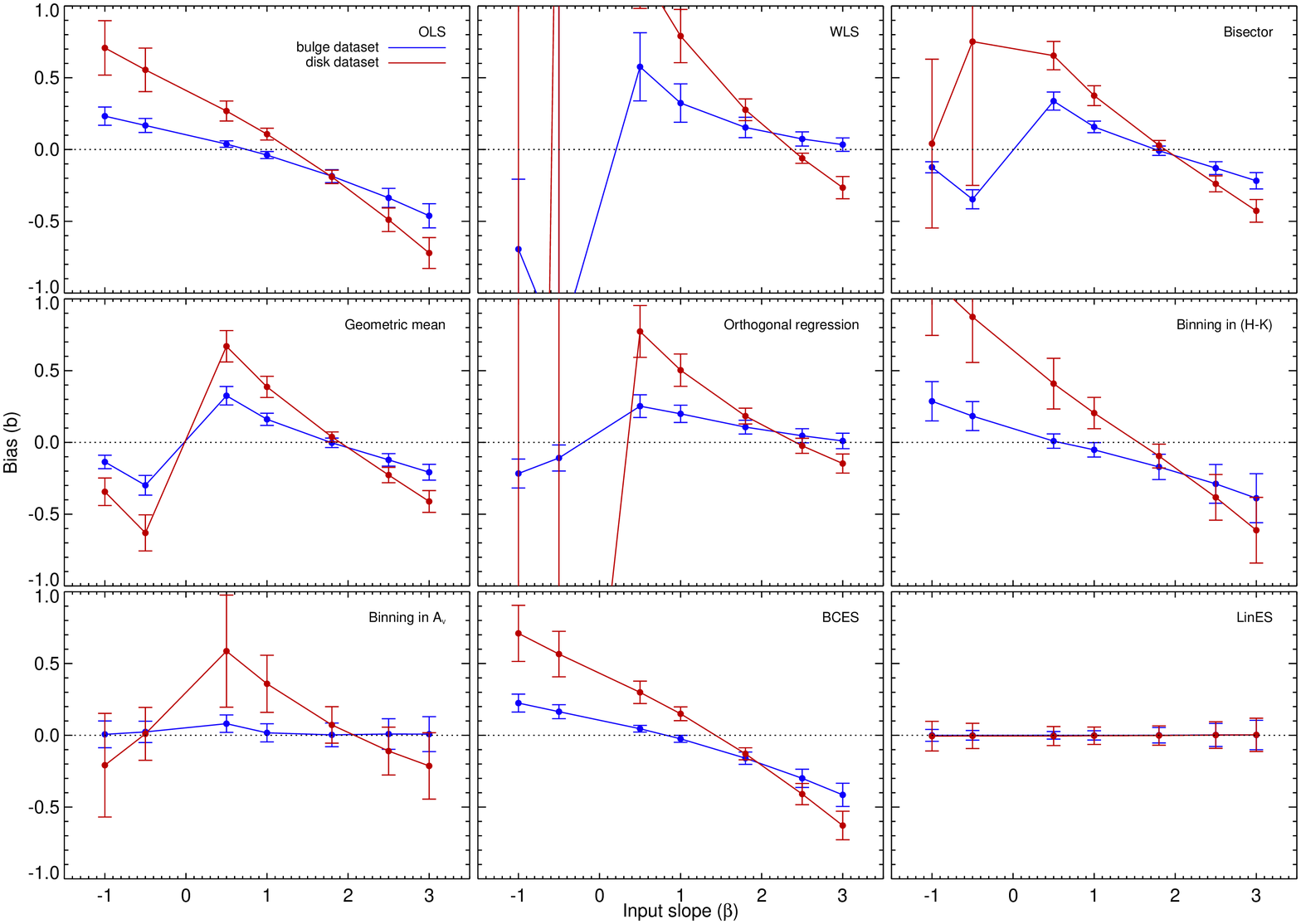}
  \caption{Bias in the slope as a function of the input slope for
    the bulge ({\it blue line}) and disk ({\it red line})
    datasets, for all the methods tested. The error bars correspond
    to the 1-$\sigma$ dispersion of the 5000 realizations.}
  \label{fig:results-set4beta}
\end{figure*}

\begin{figure*}
  \centering
  \includegraphics[width=17cm]{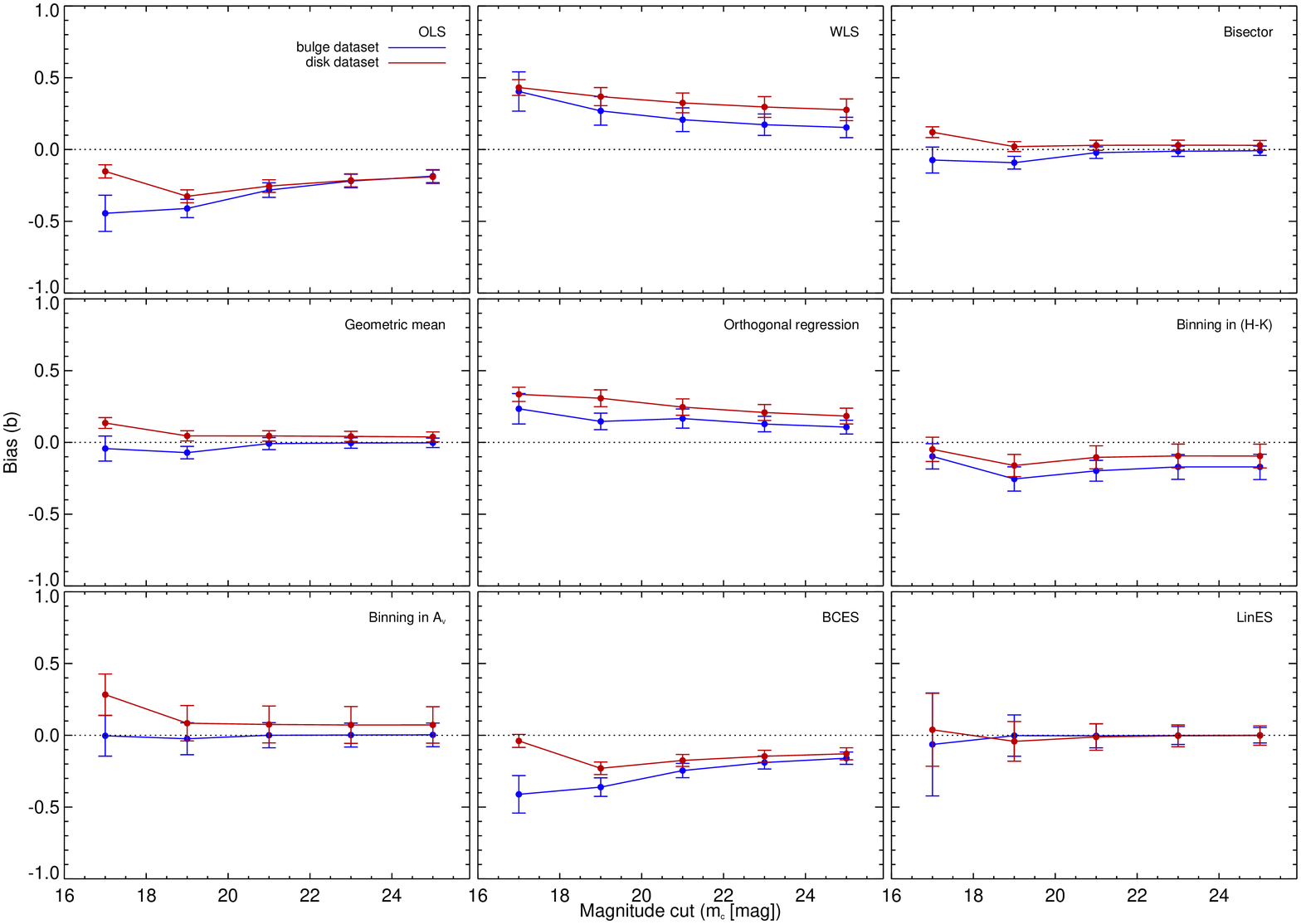}
  \caption{Bias in the slope as a function of the magnitude cut for
    the bulge ({\it blue line}) and disk ({\it red line})
    datasets, for all the methods tested. The error bars correspond
    to the 1-$\sigma$ dispersion of the 5000 realizations.}
  \label{fig:results-set4mc}
\end{figure*}

Figures \ref{fig:results-set4beta} and \ref{fig:results-set4mc} show
the bias of each method {\it vs.}  varying input slope and varying
magnitude cut, respectively, for the ``real'' datasets described in
Sec. \ref{sec:real-data}. Except for the cases described below, these
results follow those from the synthetic data, maintaining the same
behavior and changing only the magnitude of the bias.  The methods
that deserve some further consideration are the binning in $A_V$
method, the BCES, and LinES.

The BCES ({\it bottom-middle panel} in the figures) applied to these
datasets shows the same type of behavior as with the synthetic data
against varying magnitude cuts (Fig. \ref{fig:results-set4mc}). The
bias is slightly larger than before because there are fewer stars and
more dispersion. For varying input slopes, however, the bias is
considerably larger, showing that the method is severely biased for
some ranges of $\beta$ when applied to less than ideal data.

The binning in $A_V$ method, which we concluded was reasonably
unbiased when applied to the synthetic data, shows a severe bias also
for some values of the input slope $\beta$, in particular for the disk
dataset ({\it bottom-left panel in the figures}). However, this
happens because this dataset has very few stars, and a very limited
range in $A_V$ (see Figure \ref{fig:disk-dataset}), which translates
into too few $A_V$ bins to constrain the extinction vector slope
adequately. This presents a limitation of the method, away from which
it can be used with relatively good precision.

In the ``real'' datasets, LinES continues to perform
beautifully. Except for the first point in Figure
\ref{fig:results-set4mc} ({\it bottom-right panel}), where the
magnitude cut is such that limits the $A_V$ to a very narrow range,
the bias is consistent with zero for all other magnitude cuts and for
all values of the input slope we tested.

\section{Detecting a break in the extinction law}
\label{sec:detect-break-extinct}

\begin{figure}
  \resizebox{\hsize}{!}{\includegraphics{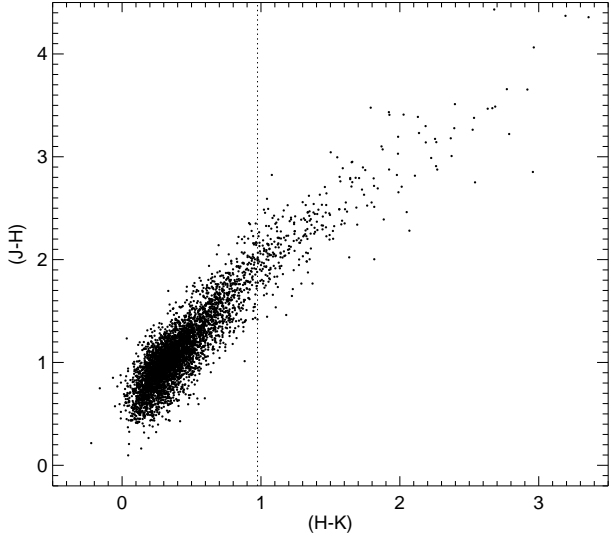}}
  \caption{Example of a color-color diagram of a synthetic stellar
    population reddened with a reddening vector with slope 1.5 up to
    $(H-K)\sim1.0$ mag, and with slope 1.0 for higher $(H-K)$.}
\label{fig:chisq_break_ccd}
\end{figure}

Grain growth at high densities has been proposed by a number of
studies (see references in Sect. \ref{sec:introduction}). If it does
occur, then in dense cores the extinction law will become grayer
toward higher extinctions, which should translate into a variation of
the slope of the reddening vector with extinction, either smooth or
abrupt depending on the nature of the transition between grain
sizes. This break was detected and measured in the Trifid Nebula for
an extinction of $A_V=20$ mag by \citet{Cambresy11}. We used the
synthetic and ``real'' datasets to test whether we could detect such a
break in the extinction law using LinES. The data was generated using
the exact same method as described above for Set 3 with $f=0.15$ and
for the ``real'' dataset (see Sect. \ref{sec:set3},
\ref{sec:real-data}), but the reddening vector was made flatter at
some value of $A_V$, or conversely, at some value of $(H-K)$ color,
since the color scales linearly with $A_V$. This produced color-color
diagrams similar to that of Fig. \ref{fig:chisq_break_ccd}, where the
break is more or less obvious depending on the difference between the
two slopes.

For each realization, we divided the sample into low-extinction
($(H-K)$ less than a value $(H-K)_\mathit{limit}$), and
high-extinction ($(H-K) > (H-K)_\mathit{limit}$), and determined the
best fits to the reddening vector in the two groups using the LinES
method, obtaining two slopes $\hat\beta_\mathit{low}$ and
$\hat\beta_\mathit{high}$.
This was done for increasing values of $(H-K)_\mathit{limit}$ in steps
of 0.2, and starting at $(H-K)_\mathit{limit} = 0.4$.

\begin{figure*}
\begin{center}
\includegraphics[width=15cm]{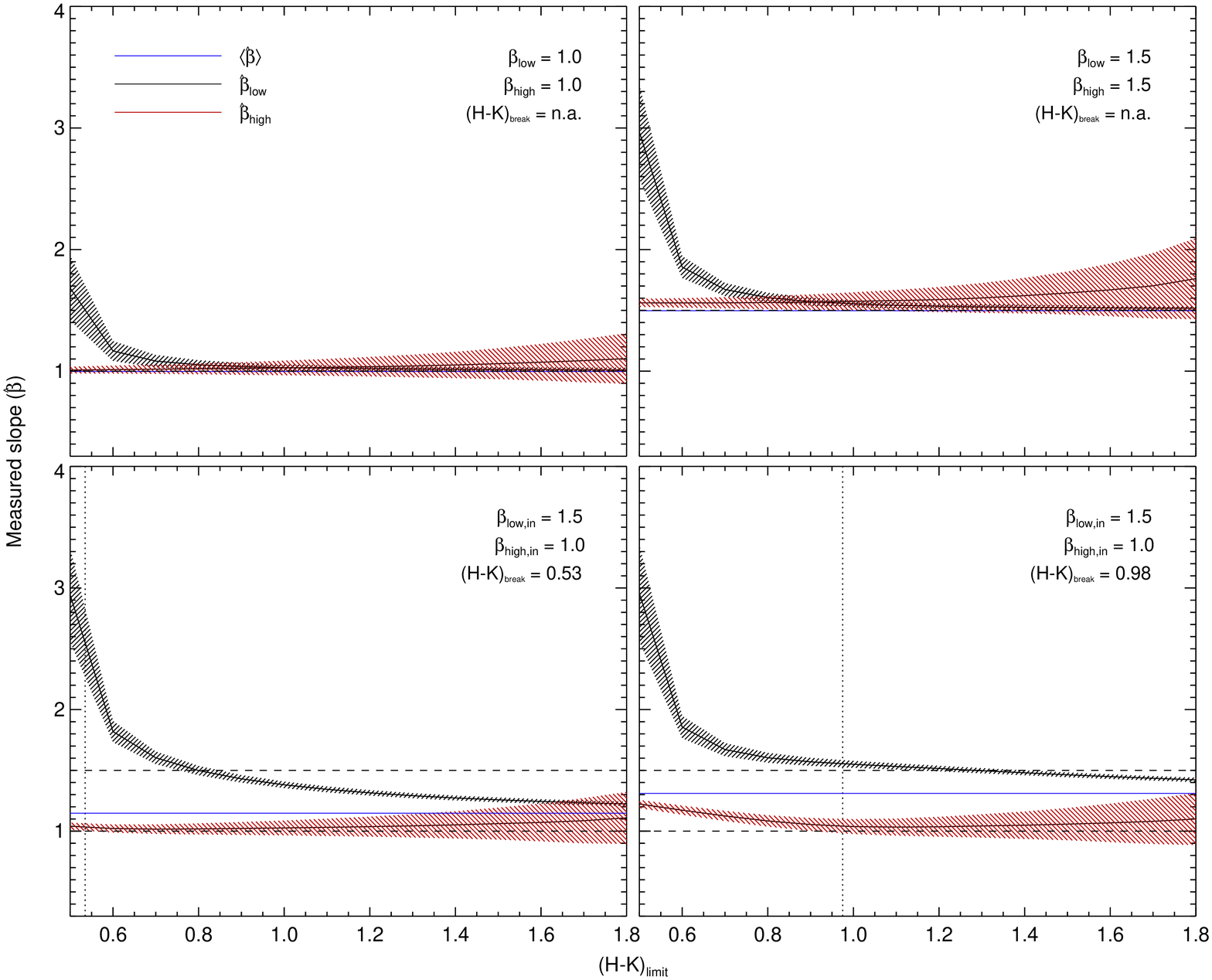}
\caption{$\hat\beta_\mathit{low}$ and $\hat\beta_\mathit{high}$ as a
  function of $(H-K)_\mathit{limit}$ for 5000 realizations of the
  synthetic dataset 3. The solid lines show the median curves, the
  shaded regions represent the $1-\sigma$ scatter from the 5000
  realizations, and the dashed lines mark the input values
  $\beta_\mathit{low}$ and $\beta_\mathit{high}$ of the slopes. {\it
    Upper panels}: $\hat\beta_\mathit{low}$ and
  $\hat\beta_\mathit{high}$ for a single-slope reddening vector with
  slopes 1.0 ({\it left}) and 1.5 ({\it right}). {\it Bottom panels}:
  same distributions but for broken extinction laws, with slopes of
  1.5 at low-extinction and 1.0 and high-extinction, and breaks
  located at $A_K=0.4$ mag ({\it left}) and $A_K=1.5$ mag ({\it
    right}). The {\it blue line} indicates the average value of the
  slope if LinES is applied to the whole range of $A_V$.}
\label{fig:slopes_break_synth}
\end{center}
\end{figure*}

\begin{figure*}
\begin{center}
\includegraphics[width=15cm]{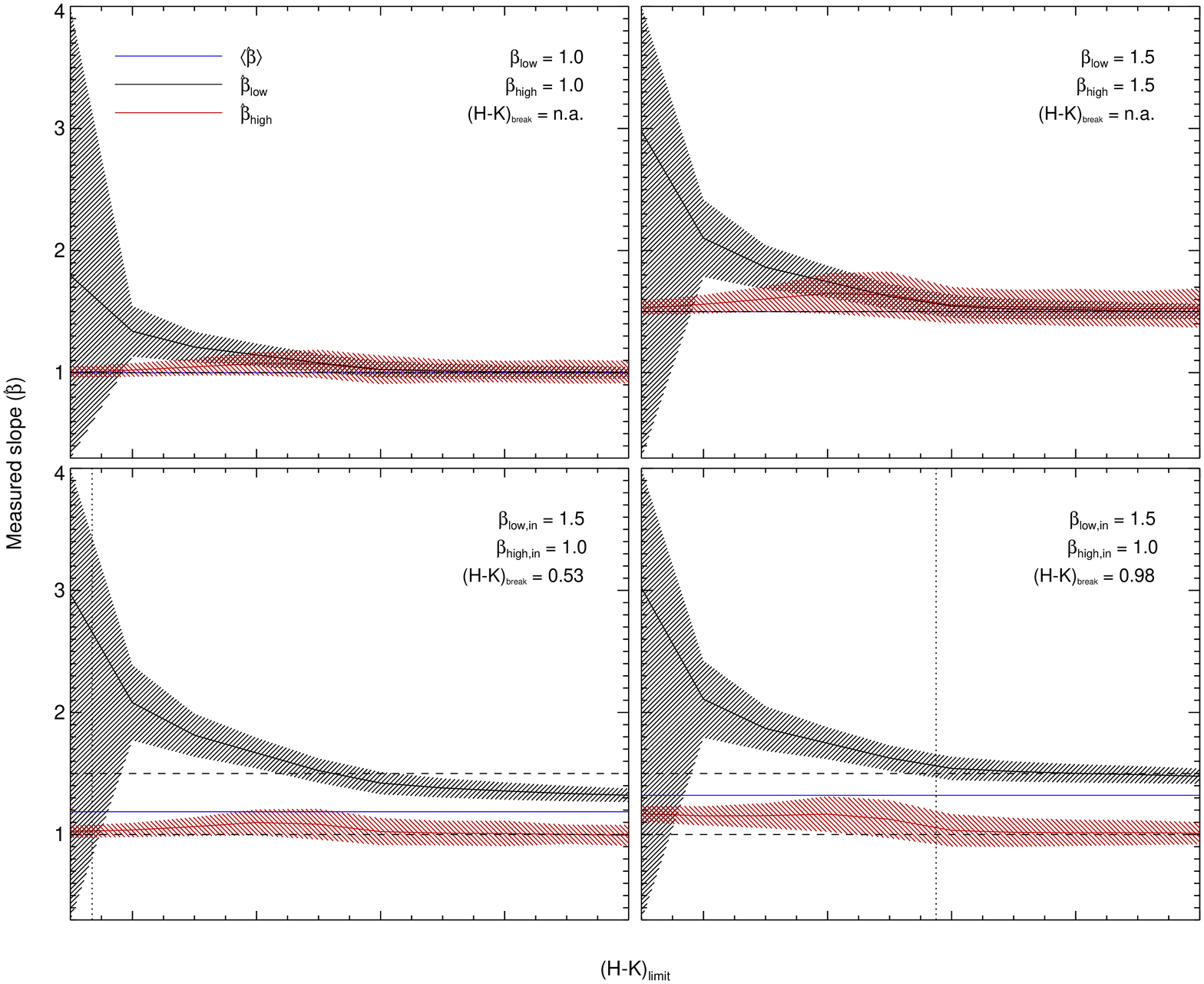}
\caption{$\hat\beta_\mathit{low}$ and $\hat\beta_\mathit{high}$ as a
  function of $(H-K)_\mathit{limit}$ for 5000 realizations of the
  bulge dataset. The solid lines show the median curves, the shaded
  regions represent the $1-\sigma$ scatter from the 5000 realizations,
  and the dashed lines mark the input values $\beta_\mathit{low}$ and
  $\beta_\mathit{high}$ of the slopes. {\it Upper panels}:
  $\hat\beta_\mathit{low}$ and $\hat\beta_\mathit{high}$ for a
  single-slope reddening vector with slopes 1.0 ({\it left}) and 1.5
  ({\it right}). {\it Bottom panels}: same distributions but for
  broken extinction laws, with slopes of 1.5 at low-extinction and 1.0
  and high-extinction, and breaks located at $A_K=0.4$ mag ({\it
    left}) and $A_K=1.5$ mag ({\it right}). The {\it blue line}
  indicates the average value of the slope if LinES is applied to the
  whole range of $A_V$.}
\label{fig:slopes_break_real}
\end{center}
\end{figure*}

Figures \ref{fig:slopes_break_synth} and \ref{fig:slopes_break_real}
show the behavior of $\hat\beta_\mathit{low}$ ({\it black}) and
$\hat\beta_\mathit{high}$ ({\it red}) as a function of
$(H-K)_\mathit{limit}$ for the synthetic dataset 3 and the ``real''
bulge dataset, respectively, when applied to 5000 realizations. The
solid lines show the median curves, and the shaded regions represent
the $1-\sigma$ scatter from the 5000 realizations. The {\it upper
  panels} are for a single-slope reddening vector with slopes 1.0
({\it left}) and 1.5 ({\it right}). The {\it bottom panels} show the
same distributions but for broken extinction laws, with slopes of 1.5
at low-extinction and 1.0 and high-extinction, and breaks located
around $A_K=0.4$ mag ({\it left}) and $A_K=1.5$ mag ({\it right}). For
completeness, we find the same results using the disk dataset, albeit
with a larger dispersion.

In the absence of a break, the two curves are indistinguishable except
for the lowest and highest values of $(H-K)_\mathit{limit}$; this is
because the subsets used for the fits in these two extremes contain
too narrow ranges in $(H-K)$ to constrain the LinES method. When a
break does exist, however, the two curves separate distinguishably,
even if the break is at low extinction. This then provides a simple
method to test whether the same extinction law applies to the full
$A_V$ range of a given dataset, or if it would rather best be
described as a two-segment law. Unfortunately this method does not
allow for the determination of the actual value of the break, but the
figures show that the slope of the extinction vector at high
extinction can be determined with reasonable accuracy. In particular,
the procedure of measuring the slopes in the low-$A_V$ and high-$A_V$
regimes provides a much better handle on the extinction law at high
extinction ({\it red line} in the figures) than measuring the slope of
the entire dataset as a whole ({\it blue line} in the figures).

\section{Summary}
\label{sec:conclusions}

We tested several methods of linear regression associated with the
problem of measuring the extinction law from photometric data. We
found that many of the commonly used methods provide biased results
caused by the presence of errors in both coordinates (which are
colors), by the fact that they are correlated, and by the presence of
scatter intrinsic to the underlying distribution.

We adapted the BCES method of \citet{Akritas:1996fk} to allow a
compensation for intrinsic scatter, using a control field to
characterize the background, unreddened population. We called this
method LinES ({\bf Lin}ear regression with {\bf E}rrors and {\bf
  S}catter). Using synthetic data, we showed this method provides
unbiased and correct results, and that it is robust against the
variation of all relevant parameters (at least) within reasonable
limits, such as size of sample, range of extinction, and amount of
intrinsic scatter.

We found that dividing any subset in sliding values of $A_V$ and
measuring the slopes of each subset can robustly differentiate between
an extinction law characterized by a single slope and one with a
break.

These results can be applied to observations of background stars seen
through dense cores of molecular clouds, or through regions that span
a reasonable range of dust density. The characterization of the
extinction law through deep, photometric data is a very useful tool to
probe the properties of the dust grains in these regions, and a
``cheap'' one when compared with, for example, spectral analysis of
many individual sources.

\begin{acknowledgements}
  We thank C. R\'oman-Z\'u\~niga for kindly providing the SOFI data,
  and the referee, L. Cambresy, for helpful comments that contributed
  to making the paper more robust.  J. Ascenso acknowledges financial
  support from FCT grant number SFRH/BPD/62983/2009. The research
  leading to these results has received funding from the European
  Community's Seventh Framework Programme (/FP7/2007-2013/) under
  grant agreement No 229517. Support for this work was also provided
  by NASA through an award issued by JPL/Caltech, contract 1279166.

\end{acknowledgements}

\bibliographystyle{aa}
\bibliography{/Users/jascenso/Dropbox/Science/bib}

\end{document}